\documentclass[preprint]{elsarticle}

\usepackage{lineno,hyperref}
\usepackage{amsmath}
\usepackage{upgreek}
\biboptions{sort&compress}
\usepackage{color}
\modulolinenumbers[5]

\journal{ArXiv}
\usepackage[labelfont={bf,sf},textfont={sf},font=scriptsize]{caption}
\usepackage[margin=0.7in]{geometry}
\usepackage{soul}

\begin{document}

\begin{frontmatter}

\title{Single chip dynamic nuclear polarization microsystem}
\author{Nergiz Sahin Solmaz}
\author{Marco Grisi}
\author{Alessandro V. Matheoud}
\author{Gabriele Gualco}
\author{Giovanni Boero\corref{mycorrespondingauthor}}
\cortext[mycorrespondingauthor]{Corresponding author}
\ead{giovanni.boero@epfl.ch}
\address{\'Ecole Polytechnique F\'ed\'erale de Lausanne (EPFL),CH-1015 Lausanne CH-1015, Switzerland}

\begin{abstract}
The integration on a single chip of the sensitivity-relevant electronics of nuclear magnetic resonance (NMR) and electron spin resonance (ESR) spectrometers is a promising approach to improve the limit of detection, especially for samples in the nanoliter and subnanoliter range. Here we demonstrate the co-integration on a single silicon chip of the front-end electronics of an NMR and an ESR detector. The excitation/detection planar spiral microcoils of the NMR and ESR detectors are concentric and interrogate the same sample volume. This combination of sensors allows to perform dynamic nuclear polarization (DNP) experiments using a single-chip integrated microsystem having an area of about 2 mm$^2$. In particular, we report $^1$H DNP-enhanced NMR experiments on liquid samples having a volume of about 1 nL performed at 10.7 GHz(ESR)/16 MHz(NMR). NMR enhancements as large as 50 are achieved on TEMPOL/H$_{2}$O solutions at room temperature. The use of state-of-the-art submicrometer integrated circuit technologies should allow the future extension of the single-chip DNP microsystem approach proposed here up the THz(ESR)/GHz(NMR) region, corresponding  the strongest static magnetic fields currently available. Particularly interesting is the possibility to create arrays of such sensors for parallel DNP-enhanced NMR spectroscopy of nanoliter and subnanoliter samples.
\end{abstract}

\begin{keyword}
\texttt{DNP}\sep \texttt{NMR}\sep \texttt{ESR}\sep \texttt{CMOS}
\end{keyword}

\end{frontmatter}


\section{Introduction}

Nuclear magnetic resonance (NMR) spectroscopy is a powerful tool employed in research, industry, and medicine. The use of NMR methodologies in an even wider range of applications is often hindered by the relatively large minimum number of resonating spins needed to achieve a sufficiently large signal-to-noise ratio (SNR) in the available experimental time.
In most of the situations studied by NMR spectroscopy the samples are concentration-limited. In these conditions, the largest SNR is obtained with the largest possible sample volume compatible with the high homogeneity region of the magnet, at the stronger possible magnetic field. However, there are also situations where the sample is volume limited. For volume limited samples, the use of an inductive detector having a sensitive volume matched to the volume of the sample under investigation  results in a significant improvement of the SNR \cite{olson1995, webb1997, lacey1999, minard2002, massin2003, sakellariou2007, maguire2007, krojanski2008, bart2009, fratila2011, zalesskiy2014, finch2016, chen2017, dupre2019}. For samples in the nanoliter and subnanoliter range, non-inductive detection methods has been also proposed, such those based on nitrogen-vacancies in diamond \cite{schirhagl2014, glenn2018, smits2019, schwartz2019, bucher2018}  and magnetic resonance force microscopy \cite{rugar1994, degen2009, mamin2007, rose2018, schnoz2019, grob2019}. Another approach to increase the SNR, applicable to samples of any volume, is to increase the nuclear spin polarization, e.g., by microwave, optical, and chemistry-based methodologies \cite{griesinger2012, slichter2014, liu2017, plainchont2018, ardenkjaer2003, capozzi2017, capozzi2015, kouvril2019, mompean2018, orlando2019, eills2019}. In the microwave dynamic nuclear polarization (DNP) approach, the sample under investigation contains unpaired electron spins which are excited into electron spin resonance (ESR). The electron spin excitation allows to enhance the nuclear magnetization of several orders of magnitude above its thermal value, improving the SNR in the NMR experiment by the same factor and reducing the required experimental time as the square of this factor. 

During the last two decades, the separate integration on a single chip of the front-end electronics of inductive NMR spectrometers \cite{boero2001, anders2009, sun2010, anders2011, anders2012, ha2014, anders2016, grisi2015, lei2016, grisi2017, sporrer2017, montinaro2018, grisi2019, handwerker2019} as well as ESR spectrometers \cite{yalcin2008, matheoud2017, anders2012b, gualco2014, matheoud2018, schlecker2018} have been demonstrated. These approaches are suitable, e.g., for the miniaturization of the probe, for the reduction of the losses and complexity of the connections, and for the realization of dense arrays of detectors. In this work we demonstrate the co-integration on a single silicon chip of the front-end electronics of an NMR and an ESR detector. The excitation/detection planar spiral microcoils of the NMR and ESR detectors are concentric and interrogate the same sample volume. This combination of sensors allows to perform dynamic nuclear polarization (DNP) experiments using a single-chip integrated DNP microsystem having an area of about 2 mm$^2$. In particular, we report $^1$H DNP experiments on liquid samples having a volume of about 1 nL performed at 10.7 GHz (ESR)/16 MHz(NMR). NMR enhancements as large as 50 are achieved on TEMPOL/H$_{2}$0 solutions at room temperature.

\section{Description of the single-chip DNP microsystem}

\begin{figure}
	\centering 
	\includegraphics[width=160mm]{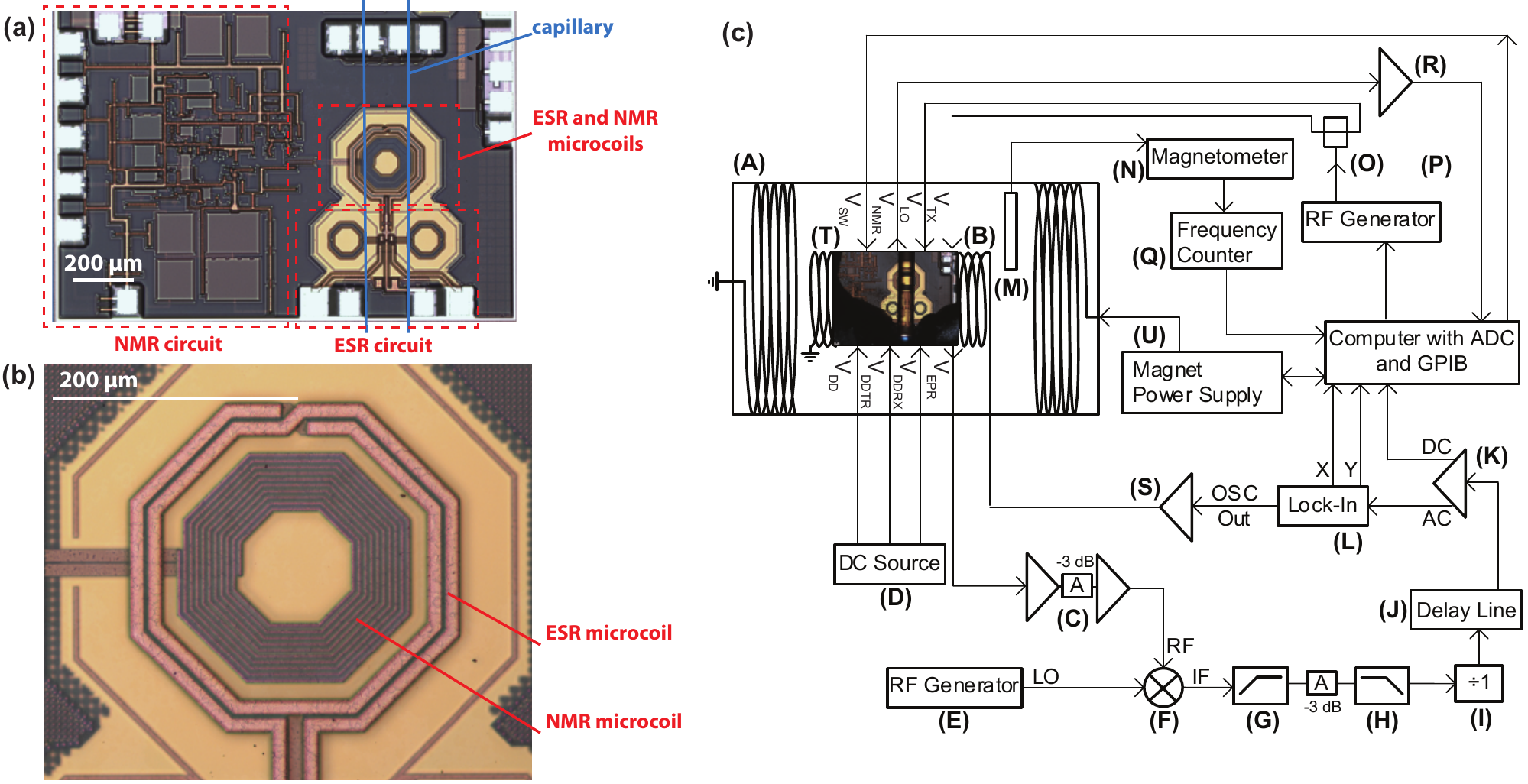} 
	\caption[]{\textbf{Set-up for the characterization of the single-chip integrated DNP microsystem operating at 10.7 GHz(ESR)/16 MHz(NMR). (a)} Photograph of the single chip DNP microsystem. The blue lines indicate the position of the capillary containing the sample under investigation. The dashed red lines indicate the NMR/ESR circuits and the concentric NMR/ESR microcoils. \textbf{(b)} Picture of the NMR and ESR microcoils. The ESR microcoil has two turns. The NMR microcoil has ten turns. \textbf{(c)} Block diagram of the complete setup for the characterization of the single-chip DNP microsystem: (A) Electromagnet (Bruker, 0 to 2.2 T); (B) Home-made modulation coil (0.33 mT/A); (C) RF amplification stage composed of two RF amplifiers (Analog Devices HMC-C001) and a 3 dB attenuator; (D) Three DC power supplies (Keithley 2400); (E) RF generator (Rohde$\&$Schwartz SMR-20) ; (F) Mixer (Mini-Circuits ZX05-153-S+); (G) 100 MHz high-pass filter (Crystek CHPFL-0100); (H) 300 MHz low-pass filter (Crystek CLPFL-0300); (I) Frequency divider (Valon Technology 3010); (J) Homemade delay-line-discriminator (200 MHz central frequency, 1 MHz detection range, 5 MHz FM bandwidth); (K) Amplifier (Stanford Research Systems SR560); (L) Lock-in amplifier (EG$\&$G 7265); (M) NMR magnetometer probe (Metrolab Instruments SA 1062 probe 3); (N) NMR magnetometer main electronic unit (Metrolab Instruments SA PT2025); (O) RF splitter (Mini-Circuits ZFSC-2-11); (P) RF generator (Stanford Research Systems SG384); (Q) Frequency counter (Fluke PM6681); (R) Amplifier (EG$\&$G 5113); (S) power amplifier (Rohrer PA508); (T) Photograph of the single-chip DNP-NMR microsystem with the capillary containing the sample placed on the microcoils and the bonding wires protected by glob-top (the black material covering the bottom part of the chip); (U) Magnet power supply (Bruker, 0 to 150 A).}  
	\label{fig:1}
\end{figure}

The single-chip DNP microsystem is composed of two parts, the ESR and NMR detectors (Fig. 1a). The ESR and NMR integrated electronic circuits are very similar to those described in details in Refs. \cite{grisi2015, matheoud2018}. The chip has a size of 1.1$\times$1.6 mm$^2$ and it is fabricated using a standard silicon CMOS technology (TSMC 180 nm, MS/RF). The ESR detector is a differential Colpitts LC oscillator operating at 10.7 GHz. The ESR microcoil has two turns and an external diameter of 270 $\mu$m, a wire width of 12 $\mu$m, a wire thickness of 2.3 $\mu$m, and a spacing of 2 $\mu$m  between turns (Fig. 1b). The NMR detector consist of a broadband (10 MHz to 1 GHz) transmit/receive electronic circuit directly connected (i.e., without tuning and matching capacitors) to an excitation/detection microcoil. The NMR microcoil has 10 turns, an external diameter of 191 $\mu$m,  a wire width of 3 $\mu$m, a wire thickness 2.3 $\mu$m, and a spacing of 2 $\mu$m between turns (Fig. 1b). With respect to the NMR microcoil used in Ref.\cite{grisi2015}, the number of turns  is reduced from 22 to 10. The reduction of the number of turns reduces the inductance from 190 nH to 17 nH and the series resistance from 190 $\Omega$ to 35 $\Omega$. In these conditions, the RF current and the dissipated power in the NMR microcoil during excitation at 16 MHz is of 90 mA and 280 mW, respectively. The NMR microcoil is concentrically placed inside the ESR microcoil (Fig.1b). The two coupled concentric microcoils are simulated using a full wave electromagnetic simulator (Advanced Design System ADS, Keysight Technologies). The obtained S-parameter files are used in an integrated circuits simulator (Cadence, Cadence Design Systems Inc.) to simulate both the ESR and the NMR integrated detectors. In these simulations it is found that an NMR coil with more than 15 turns significantly reduces the quality factor of the ESR coil and quenches the oscillation of the ESR circuitry. This is qualitatively explained by the losses caused by the induced currents in the NMR coil. To keep a safety margin, the implemented NMR coil has 10 turns. The DC power consumption of the NMR system is 70 mW in the transmit mode and 40 mW in the receive mode. The ESR detector is a slightly modified version of the one reported in Ref.\cite{matheoud2018}, where the bonding pads are distanced to facilitate the placement of a capillary containing the liquid sample under investigation (Fig. 1a,c). The DC power consumption of the ESR circuit is 90 mW at the maximum supply voltage of the oscillator of 2 V, and 5 mW at the minimum supply voltage of the oscillator of 0.85 V.

\begin{figure*}
	\centering 
	\includegraphics[width=160mm]{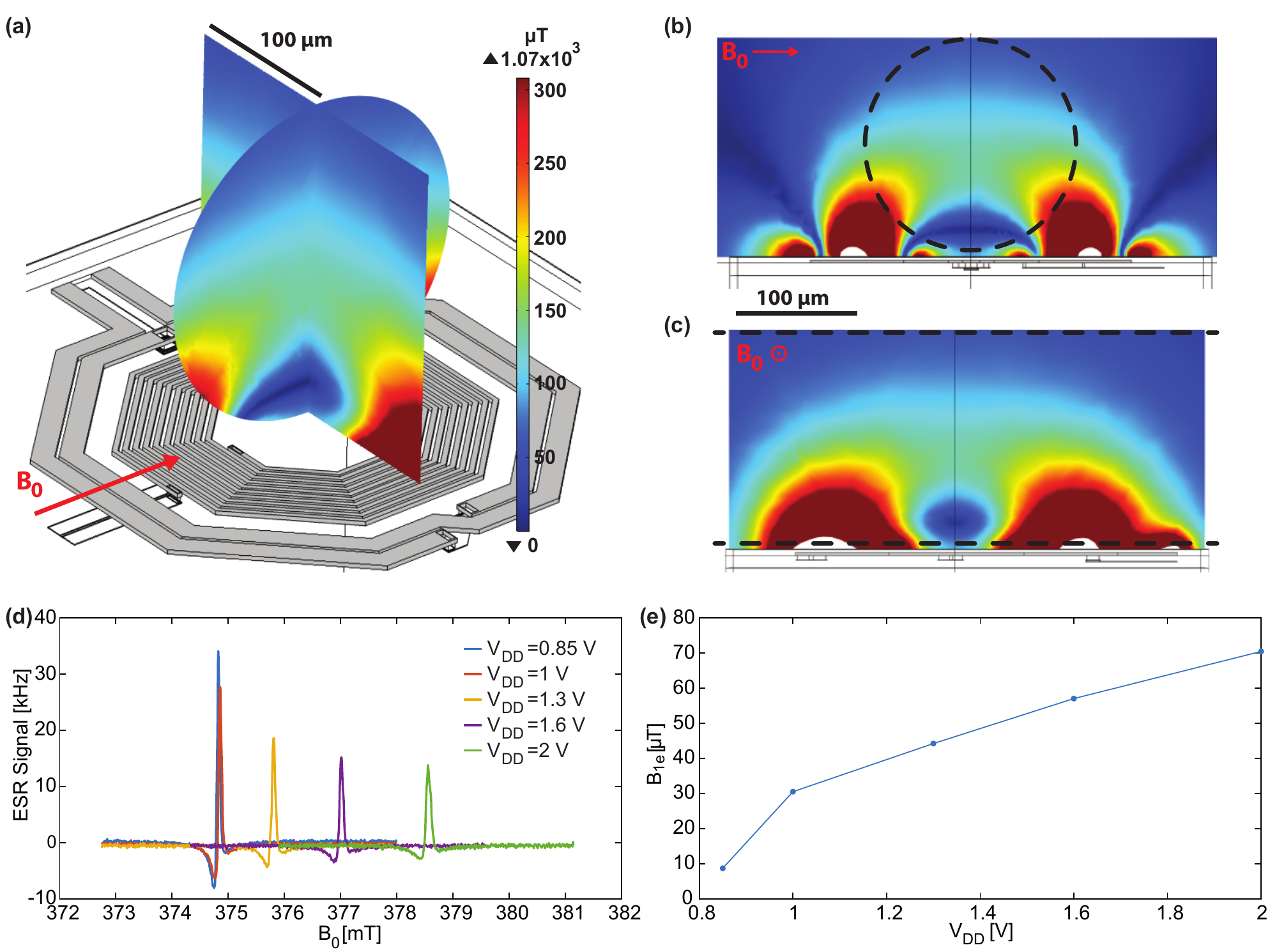} 
	\caption[]{\textbf{Simulations and experiments for the characterization of the microwave magnetic field produced by the ESR microcoil.} \textbf{(a)} Three dimensional representation of the ESR and NMR microcoils together with the map of the microwave magnetic field $B_{1e}$, defined as half of the component perpendicular to the static magnetic field $B_0$ (indicated by the red arrow) of the microwave magnetic field produced by the microwave current into ESR microcoil. In \textbf{(a)}, the two perpendicular cross-sections corresponds to region occupied by the water solution inside the capillary. In \textbf{(b)} and \textbf{(c)} are reported the maps of $B_{1e}$ in two larger regions. The black dashed lines indicate the region occupied by the water solution inside the capillary. The simulation are performed using COMSOL Multiphysics (COMSOL Inc.). The amplitude of the microwave current in the ESR microcoil is set to 74 mA, according to the combined results of simulations of the ESR/NMR integrated electronics performed with Advanced Design System (ADS, Keysight Technologies) and Cadence (Cadence Design Systems Inc.). \textbf{(d)} ESR spectra of a sample of BDPA placed in the center of the ESR microcoil for different ESR oscillator supply voltages $V_{DD}$. The increase of the resonance static magnetic field is due to the increase of the oscillator frequency with the oscillator supply voltage $V_{DD}$. \textbf{(e)} Microwave magnetic field $B_{1e}$ in the center of the ESR coil extracted from the measurement of the linewidth of the ESR signals shown in \textbf{(d)} according to the equation $\Delta {B_{0,zc}} = (2/\gamma {T_2}){(1 + {\gamma ^2}B_1^2{T_1}{T_2})^{1/2}}$ \cite{poole1996}, where $\Delta {B_{0,zc}}$ is the field difference between the two zero-crossing of the ESR signal and it corresponds to the peak-to-peak linewidth of the dispersion signal measurable without magnetic field modulation. Experimental conditions: $f_{mw}\cong$ 10.7 GHz, modulation frequency: $f_m=$ 16.7 kHz, modulation magnetic field: $B_m=$ 6 $\mu$T.}
	\label{fig:Byz}
\end{figure*}

The complete setup for the characterization of the single chip DNP microsystem is shown in Fig.\ref{fig:1}c. The frequency-to-voltage conversion of the ESR detector output is performed by a delay-line-discriminator (DLD) whose central frequency is 200 MHz. In order to match the DLD central frequency and improve the spectral purity, the signal at the output of the detector is amplified, mixed with an external reference, filtered, and shaped through a divide-by-1 frequency divider. The DC coupled output of the DLD is amplified and digitized by an analog-to-digital converter (ADC). This signal is used to monitor the ESR oscillator frequency. The AC coupled output of the DLD is amplified and sent to a lock-in amplifier for further amplification and synchronous demodulation. The lock-in amplifier also generates the reference signal which is used for the magnetic field modulation. An NMR magnetometer is used to track the variations of the magnetic field produced by the electromagnet in which the DNP experiments are performed. The frequency-lock with the NMR magnetometer is necessary due to the relatively large drift (about 1 ppm/h) of the electromagnet. The NMR frequency measured by the magnetometer is used to set the frequency of the RF generator connected to integrated transmitter $V_{TX}$ and to the local oscillator $V_{LO}$ of the integrated receiver. In particular, the RF generator frequency is set 20 kHz below the frequency measured by the NMR magnetometer. This allows to obtain an NMR signal above the $1/f$-noise corner frequency of the integrated NMR receiver. A switch signal $V_{SW}$ determines the pulse width of the NMR excitation. The output of the integrated NMR receiver $V_{NMR}$ is amplified, digitized, and digitally processed. The maximum NMR signal amplitude is obtained with a radiofrequency pulse length $\tau_{rf}\cong$ 5 $\mu$s. The maximum microwave magnetic field $B_{1e}$ at the center of the ESR coil is about 70 $\mu$T, as estimated from measurements performed on a BDPA sample (see below).

The samples are contained into borosilicate glass capillaries (BGCT 0.2, Capillary Tube Supply Ltd)  of 0.2 mm outer diameter (OD) and 0.18 mm inner diameter (ID). The capillary are sealed with a torch (Microtorch, Prodont Holliger). The non-degassed solutions of pure water (H$_{2}$O)(W3500, Sigma-Aldrich) and  4-hydroxy-2,2,6,6-tetramethylpiperidine 1-oxyl (TEMPOL)(176141, Sigma-Aldrich) are obtained by dilution at room temperature in air starting from a 1 M solution. The degassed solutions are prepared as follows. The degassing of water is performed by bubbling with a nitrogen flow for about 1 h. A 200 mL glass vial, with inlet/outlet pipes for the nitrogen flow, is filled with pure water heated to 70 $^{\circ}$C by a hot plate. A pipette is used to transfer the degassed water into a 2 mL glass vial containing the appropriate amount of TEMPOL molecules to produce a 100 mM solution. The 10 mM and 1 mM solutions are prepared into 2 mL glass vials by subsequent dilution in degassed water. The preparation of the solutions and the filling/sealing of the capillaries is performed in a few seconds to minimize the absorption of O$_2$ in contact with air (the diffusion length of O$_2$ in water in 1 s is about 60 $\mu$m). 

Electromagnetic and electrical simulation of the integrated microwave oscillator, performed with ADS and Cadence, show that the maximum achievable microwave current $I_{mw}$ in the ESR microcoil is of 74 mA, obtained with an oscillator supply voltage of 2 V. Figure \ref{fig:Byz} shows the result of a COMSOL simulation performed with microwave current $I_{mw}=$74 mA in the ESR microcoil. The indicated magnetic field is half of the magnitude of the component of the microwave magnetic field perpendicular to the static magnetic field (i.e., $B_{1e}$) in two orthogonal cross-sections of the capillary where the sample is confined. The microwave magnetic field $B_{1e}$ in the center of the coil is about 100 $\mu$T. The obtained microwave magnetic field is the result of the superposition of the microwave magnetic field created by the ESR coil and the  microwave magnetic field caused by the induced microwave currents in the NMR microcoil. Moving vertically from the chip (and coils) surface into the sample region, the total microwave magnetic field first decreases, cancels out in the dark blue region, then increases,  and finally decrease again further away from the chip surface. This is the expected magnetic field produced by two concentric coils of different diameter carrying currents flowing in opposite directions. Simulations of the microwave magnetic field produced by the ESR microcoil with the NMR microcoil terminated with a high impedance (or without the NMR microcoil) show that the average microwave magnetic field in the sample region could be increased by almost one order of magnitude (and, as expected, in the center of the coil would be approximately given by $B_{1e}$ = $\mu_0 I_{mw}/d\cong$ 400 $\mu$T).

To cross-check these simulation results, we performed experiments with a single crystal of 1:1$\alpha$,$\gamma$-bisdiphenylene-$\beta$-phenylallyl (BDPA/benzene, 152560, Sigma-Aldrich) having a size of about 70$\times$70$\times$5 $\mu$m$^3$ placed in the center of the ESR microcoil (Fig. \ref{fig:Byz}d). At room temperature, BDPA has relaxation times $T_1\cong T_2 \cong $ 100 ns \cite{goldsborough1960}. Since the linewidth depends only on $B_{1e}$ and the relaxation times \cite{poole1996}, the knowledge of the relaxation times and the measurement of the linewidth allows to estimate value of $B_{1e}$. From the measured ESR signals reported in Fig. \ref{fig:Byz}d, the extracted value of $B_{1e}$ in the center of the ESR microcoil is about 70 $\mu$T at the maximum supply voltage of the oscillator of 2 V (Fig. \ref{fig:Byz}e), corresponding to the simulated maximum current in the ESR coil of 74 mA. As discussed above, the full wave COMSOL simulation gives a $B_{1e}$ in the center of the ESR microcoil of about 100 $\mu$T, in good agreement with the 70 $\mu$T estimated from these BDPA measurements. 

\section{Experimental results}

\begin{figure}
	\centering 
	\includegraphics[width=160mm]{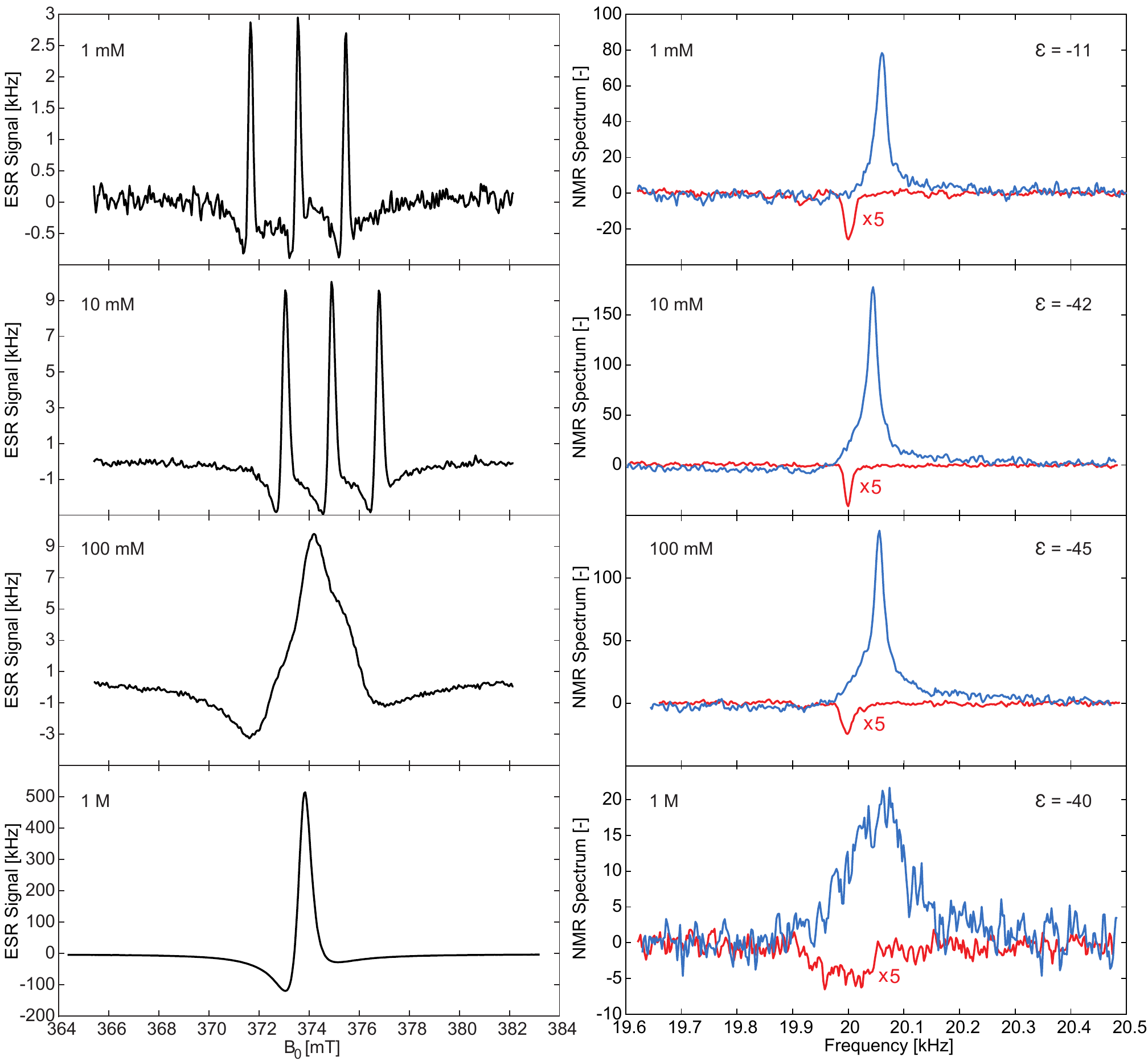} 
	\caption[]{\textbf{ESR (left column) and NMR (right column) spectra of TEMPOL/H$_2$O solutions for different concentrations (1mM , 10 mM, 100 mM, and 1 M).} In the NMR spectra, the red curves are the non-enhanced ($B_{1e}=0$) NMR spectra enlarged 5 times whereas the blue curves are the DNP-enhanced NMR spectra ($B_{1e}$$\cong$ 60 $\mu$T). The DNP enhancement $\varepsilon$ values for each spectra is given on the top-right corner. The enhancement is defined as the ratio of the integrals of the enhanced and non-enhanced NMR signals in the frequency domain. The ESR measurements are performed in the following conditions: modulation frequency $f_m=16.7$ kHz, modulation magnetic field $B_m\cong$ 6 $\mu$T, microwave frequency:$f_{mw}\cong$ 10.7 GHz, microwave magnetic field: $B_{1e}$$\cong$ 60 $\mu$T. The NMR measurements are performed in the following conditions: $f_{rf}\cong$ 16 MHz, pulse length $\tau_{rf}=5~\mu s$, pulse repetition time  $T_r=500$ ms, time-domain match filter time constant $T_m=100$ ms, acquisition time $T_{daq}=400$ ms, number of averaging $N_{avg}=100000$ (for the non-enhanced signal) and $N_{avg}=1000$ for the enhanced signal.}
	\label{fig:ESR_NMR_DNP}
\end{figure}

The capillary encapsulated samples of TEMPOL/H$_{2}$O solutions are fixed on top of the single-chip DNP microsystem with a small drop of vacuum grease (high vacuum grease, Dow Corning). As shown in the left side of Fig.\ref{fig:ESR_NMR_DNP}, at concentrations of 1 mM and 10 mM the ESR spectra consist of three hyperfine lines due the $^{15}$N nucleus ($I=1$). At concentrations of 100 mM and 1 M a single line is observed, as reported also in Ref. \cite{gafurov2013}. After obtaining the ESR spectrum, the $B_0$ magnetic field is set to one of the three maxima for concentrations of 1 mM and 10 mM and to the single maximum for the 100 mM and 1 M concentrations. For all NMR spectra shown in the right side of Fig. 2, the RF pulse length is $\tau_{rf}=$ 5 $\mu$s, the acquisition time is $T_{daq}=$ 400 ms, and the pulse repetition time is $T_{r}$= 500 ms. The non-enhanced NMR spectra are the average of $N_{avg}=$ 100000 spectra obtained in about 14 hours. The DNP-enhanced NMR spectra are the average of $N_{avg}=$ 1000 spectra obtained in about 9 minutes. In the DNP-enhanced NMR measurements, the microwave excitation is present also during the NMR detection. The frequency shift between the non-enhanced and enhanced NMR spectra is caused by the DC current flowing in the two turn ESR microcoil. Even though the coil is designed such that the magnetic field created by the DC current running in each section is almost entirely canceled by the magnetic field created by the DC current running in the adjacent section, a small field is still created in the direction of $B_0$ due to the non-zero distance (about 14 $\mu$m) between the two wires. The observed shift (up to 4 ppm) is much larger than the one attributable to temperature effects. Since the temperature induced frequency shift for the $^1$H nucleus in water is about 0.01 ppm/$^{\circ}$C (see Ref.\cite{petley1984}), the observed shift would correspond to a temperature increase of 400 $^{\circ}$C, which is obviously impossible for a liquid water sample. In a future version of the integrated ESR detector, this shift will be entirely suppressed by an ESR oscillator design in which no DC current runs through the ESR microcoil, such as the Colpitts oscillator reported in Ref. \cite{matheoud2018}.  

\begin{figure*}
	\centering 
	\includegraphics[width=160mm]{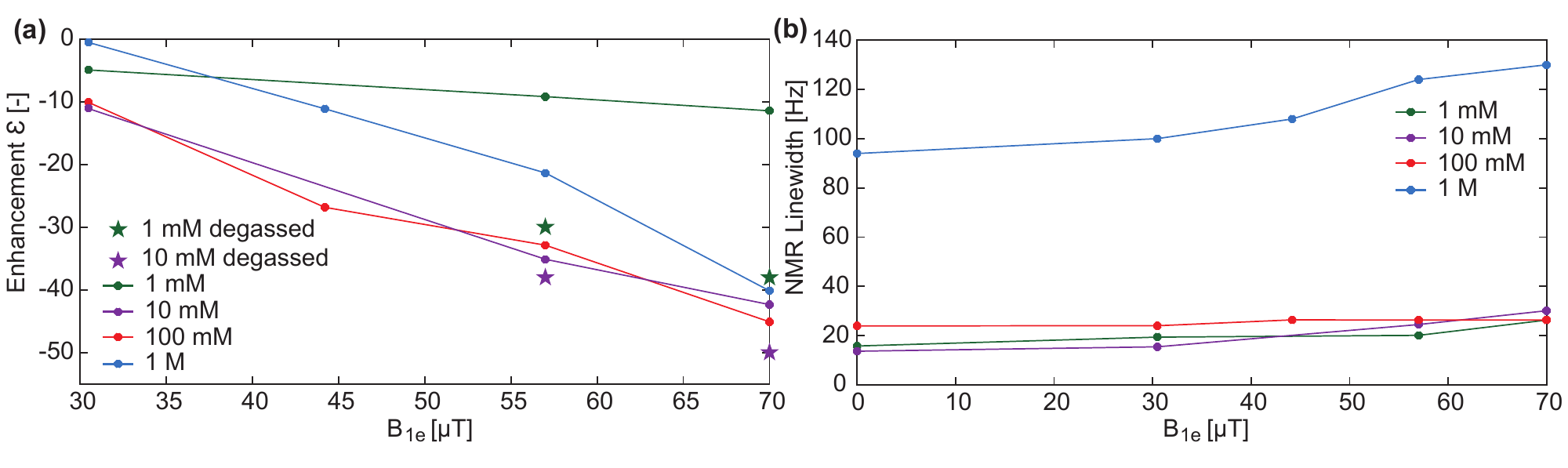} 
	\caption[]{\textbf{DNP enhancement and NMR linewidth. (a)} Enhancement $\varepsilon$ and \textbf{(b)} NMR linewidth  of TEMPOL/H$_2$O solutions with different concentrations (1mM, 10 mM, 100 mM, and 1 M) at different microwave magnetic field $B_{1e}$. The enhancement is defined as the ratio of the integrals of the enhanced and non-enhanced NMR signals in the frequency domain. The NMR linewidth is defined as the full width at half maximum of the NMR signal in the frequency domain. At the operating frequency of 16 MHz, a linewidth of 20 Hz corresponds to 1.25 ppm.}
	\label{fig:Enh_LW}
\end{figure*}

Fig.\ref{fig:Enh_LW} reports the DNP enhancements (Fig.\ref{fig:Enh_LW}a) and the NMR linewidths (Fig.\ref{fig:Enh_LW}b) obtained with TEMPOL/H$_2$O solutions having different concentrations. A maximum enhancement of about $-$50 is obtained with the largest $B_{1e}$ (i.e., about 70 $\mu$T) with a 10 mM degassed solution. Enhancements in the order of $-$100 have been previously reported at 0.3 T \cite{hofer2008,hofer2008b, mccarney2007,turke2010, enkin2014, turke2012, munnemann2008}. In particular, in Ref. \cite{hofer2008}, an enhancement of $-$100 is reported for a 10 mM solution of TEMPOL/H$_2$O. The lower enhancement measured in our work is probably due to the lower average $B_{1e}$ which, as shown in Fig.\ref{fig:Enh_LW}a, is insufficient to reach the saturation region. This is caused by the NMR coil inside the ESR coil, which reduces significantly the microwave magnetic field $B_{1e}$ produced by the ESR coil as explained in the description of the system. Figure \ref{fig:Enh_LW} shows also that the degassing of water increases the enhancement, especially for low radical concentrations. The enhancement of the NMR signal for the 1 mM degassed solution is almost three times larger than the one measured with the non-degassed solution. For the 10 mM sample, the enhancement increase caused by the degassing is much lower, probably due to the negligible additional relaxation induced by the presence of the O$_2$ molecules (0.2 mM at room temperature) with respect to the relaxation due to 10 mM TEMPOL molecules. A shown in Fig.\ref{fig:Enh_LW}b, for 1 mM to 100 mM concentrations, the NMR linewidth is about 20 Hz (i.e., 1 ppm), mainly limited by the lack of shim coils in the electromagnet (a very similar chip operated in a 7 T magnet equipped with shimming coils shows spectral resolutions down to about 2 Hz \cite{montinaro2018}).  The NMR linewidth of the 1 M solution is about 90 Hz, presumably due to the large concentration of TEMPOL which significantly reduces the $T_{2}$ value \cite{asakura1991, hofer2008}. 

\section{Outlook}
In this work we demonstrated, for the first time, the integration on a single silicon chip of less than 2 mm$^2$ of a DNP microsytem consisting of an NMR transceiver and an ESR oscillator. Measurements on TEMPOL/H$_2$O solutions, performed at 10.7 GHz(ESR)/16 MHz(NMR), show enhancements as large as $-$50 on samples having an effective volume of about 1 nL. In the following, we discuss the foreseen improvements and extensions of the approach demonstrated in this work. A straightforward but rather modest increase of the signal-to-noise ratio can be obtained by a narrowband design of the integrated NMR receiver. A very significant improvement of the signal-to-noise ratio could be obtained by increasing the operating frequency of the NMR/ESR subsystems. A moderate increase of the frequency to the 40 GHz(ESR)/60 MHz(NMR) region could allow for low-cost DNP-enhanced NMR spectrometers in permanent magnets. The use of state-of-the-art submicrometer integrated circuit technologies should allow the extension of the single-chip DNP microsystem approach proposed here up the THz(ESR)/GHz(NMR) region \cite{razavi2011,takahashi2017,razavi2012}, corresponding  the strongest static magnetic fields currently available. In Ref.\cite{matheoud2017} we reported about single-chip ESR detectors operating up to 146 GHz and in Refs.\cite{grisi2015, grisi2017, grisi2019} about single-chip NMR detectors operating up to 300 MHz. However, the combination of NMR/ESR detectors at these frequencies (and above) has been not yet demonstrated. The main technical challenge for the extension of the proposed single-chip DNP microsystem approach to higher frequencies is the coupling between the NMR and ESR excitation/detection structures, which influence the strength of the microwave field $B_{1e}$. In order to obtain a sufficiently large $B_{1e}$, the NMR structure and its impedance termination should be carefully co-designed with the ESR structure, with possible drawbacks in terms of NMR sensitivity. Another interesting opportunity offered by the single-chip approach is the possibility to create dense arrays of such sensors for parallel DNP-enhanced NMR spectroscopy of a large number of nanoliter and subnanoliter different samples (or a bigger volume of the same sample). Additionally, preliminary measurements performed with the DNP microsystem proposed in this work show that it can be operated also at temperatures down to 4 K, at least. Hence, the single-chip DNP approach proposed here could be well suited also for the study of DNP processes other then the Overhauser effect in liquids at room temperature. 

\section{Acknowledgments}
Financial support from the Swiss National Science Foundation (SNSF) is gratefully acknowledged (grant 200020-175939).
We thank Giancarlo Corradini for wire bonding and bonding protection, Pierrick Clement for his helps in capillary sealing, and Mohammadmahdi Kiaee for his help in the degassing of the samples.
 
\bibliography{mybibfile}

\begin{thebibliography}{10}
\expandafter\ifx\csname url\endcsname\relax
  \def\url#1{\texttt{#1}}\fi
\expandafter\ifx\csname urlprefix\endcsname\relax\def\urlprefix{URL }\fi
\expandafter\ifx\csname href\endcsname\relax
  \def\href#1#2{#2} \def\path#1{#1}\fi

\bibitem{olson1995}
D.~L. Olson, T.~L. Peck, A.~G. Webb, R.~L. Magin, J.~V. Sweedler,
  High-resolution microcoil $^1${H}-{NMR} for mass-limited, nanoliter-volume
  samples, Science 270~(5244) (1995) 1967--1970.

\bibitem{webb1997}
A.~G. Webb, Radiofrequency microcoils in magnetic resonance, Progress in
  Nuclear Magnetic Resonance Spectroscopy 31~(1) (1997) 1--42.

\bibitem{lacey1999}
M.~E. Lacey, R.~Subramanian, D.~L. Olson, A.~G. Webb, J.~V. Sweedler,
  High-resolution {NMR} spectroscopy of sample volumes from 1 nl to 10 $\mu$l,
  Chemical reviews 99~(10) (1999) 3133--3152.

\bibitem{minard2002}
K.~R. Minard, R.~A. Wind, Picoliter $^1$h {NMR} spectroscopy, Journal of
  Magnetic Resonance 154~(2) (2002) 336--343.

\bibitem{massin2003}
C.~Massin, F.~Vincent, A.~Homsy, K.~Ehrmann, G.~Boero, P.-A. Besse, A.~Daridon,
  E.~Verpoorte, N.~De~Rooij, R.~Popovic, Planar microcoil-based microfluidic
  {NMR} probes, Journal of Magnetic Resonance 164~(2) (2003) 242--255.

\bibitem{sakellariou2007}
D.~Sakellariou, G.~Le~Goff, J.-F. Jacquinot, High-resolution, high-sensitivity
  {NMR} of nanolitre anisotropic samples by coil spinning, Nature 447~(7145)
  (2007) 694--697.

\bibitem{maguire2007}
Y.~Maguire, I.~L. Chuang, S.~Zhang, N.~Gershenfeld, Ultra-small-sample
  molecular structure detection using microslot waveguide nuclear spin
  resonance, Proceedings of the National Academy of Sciences 104~(22) (2007)
  9198--9203.

\bibitem{krojanski2008}
H.~G. Krojanski, J.~Lambert, Y.~Gerikalan, D.~Suter, R.~Hergenr{\"o}der,
  Microslot {NMR} probe for metabolomics studies, Analytical chemistry 80~(22)
  (2008) 8668--8672.

\bibitem{bart2009}
J.~Bart, A.~J. Kolkman, A.~J. Oosthoek-de Vries, K.~Koch, P.~J. Nieuwland,
  H.~Janssen, J.~van Bentum, K.~A. Ampt, F.~P. Rutjes, S.~S. Wijmenga, et~al.,
  A microfluidic high-resolution {NMR} flow probe, Journal of the American
  Chemical Society 131~(14) (2009) 5014--5015.

\bibitem{fratila2011}
R.~M. Fratila, A.~H. Velders, Small-volume nuclear magnetic resonance
  spectroscopy, Annual review of analytical chemistry 4 (2011) 227--249.

\bibitem{zalesskiy2014}
S.~S. Zalesskiy, E.~Danieli, B.~Blumich, V.~P. Ananikov, Miniaturization of
  {NMR} systems: Desktop spectrometers, microcoil spectroscopy, and “{NMR} on
  a chip” for chemistry, biochemistry, and industry, Chemical reviews
  114~(11) (2014) 5641--5694.

\bibitem{finch2016}
G.~Finch, A.~Yilmaz, M.~Utz, An optimised detector for in-situ high-resolution
  {NMR} in microfluidic devices, Journal of Magnetic Resonance 262 (2016)
  73--80.

\bibitem{chen2017}
Y.~Chen, H.~S. Mehta, M.~C. Butler, E.~D. Walter, P.~N. Reardon, R.~S. Renslow,
  K.~T. Mueller, N.~M. Washton, High-resolution microstrip {NMR} detectors for
  subnanoliter samples, Physical Chemistry Chemical Physics 19~(41) (2017)
  28163--28174.

\bibitem{dupre2019}
A.~Dupr{\'e}, K.-M. Lei, P.-I. Mak, R.~P. Martins, W.~K. Peng, Micro-and
  nanofabrication {NMR} technologies for point-of-care medical applications--a
  review, Microelectronic Engineering (2019).

\bibitem{schirhagl2014}
R.~Schirhagl, K.~Chang, M.~Loretz, C.~L. Degen, Nitrogen-vacancy centers in
  diamond: nanoscale sensors for physics and biology, Annual review of physical
  chemistry 65 (2014) 83--105.

\bibitem{glenn2018}
D.~R. Glenn, D.~B. Bucher, J.~Lee, M.~D. Lukin, H.~Park, R.~L. Walsworth,
  High-resolution magnetic resonance spectroscopy using a solid-state spin
  sensor, Nature 555~(7696) (2018) 351--354.

\bibitem{smits2019}
J.~Smits, J.~T. Damron, P.~Kehayias, A.~F. McDowell, N.~Mosavian, I.~Fescenko,
  N.~Ristoff, A.~Laraoui, A.~Jarmola, V.~M. Acosta, Two-dimensional nuclear
  magnetic resonance spectroscopy with a microfluidic diamond quantum sensor,
  Science Advances 5~(7) (2019).

\bibitem{schwartz2019}
I.~Schwartz, J.~Rosskopf, S.~Schmitt, B.~Tratzmiller, Q.~Chen, L.~P.
  McGuinness, F.~Jelezko, M.~B. Plenio, Blueprint for nanoscale {NMR},
  Scientific reports 9~(1) (2019) 6938.

\bibitem{bucher2018}
D.~B. Bucher, D.~R. Glenn, H.~Park, M.~D. Lukin, R.~L. Walsworth,
  Hyperpolarization-enhanced {NMR} spectroscopy with femtomole sensitivity
  using quantum defects in diamond, arXiv preprint arXiv:1810.02408 (2018).

\bibitem{rugar1994}
D.~Rugar, O.~Z{\"u}ger, S.~Hoen, C.~S. Yannoni, H.-M. Vieth, R.~D. Kendrick,
  Force detection of nuclear magnetic resonance, Science 264~(5165) (1994)
  1560--1563.

\bibitem{degen2009}
C.~Degen, M.~Poggio, H.~Mamin, C.~Rettner, D.~Rugar, Nanoscale magnetic
  resonance imaging, Proceedings of the National Academy of Sciences 106~(5)
  (2009) 1313--1317.

\bibitem{mamin2007}
H.~Mamin, M.~Poggio, C.~Degen, D.~Rugar, Nuclear magnetic resonance imaging
  with 90-nm resolution, Nature nanotechnology 2~(5) (2007) 301.

\bibitem{rose2018}
W.~Rose, H.~Haas, A.~Q. Chen, N.~Jeon, L.~J. Lauhon, D.~G. Cory, R.~Budakian,
  High-resolution nanoscale solid-state nuclear magnetic resonance
  spectroscopy, Physical Review X 8~(1) (2018) 011030.

\bibitem{schnoz2019}
S.~Schnoz, A.~D{\"a}pp, A.~Hunkeler, B.~H. Meier, Detection of liquids by
  magnetic resonance force microscopy in the gradient-on-cantilever geometry,
  Journal of Magnetic Resonance 298 (2019) 85--90.

\bibitem{grob2019}
U.~Grob, M.-D. Krass, M.~Heritier, R.~Pachlatko, J.~Rhensius, J.~Kosata,
  B.~Moores, H.~Takahashi, A.~Eichler, C.~L. Degen, Magnetic resonance force
  microscopy with a one-dimensional resolution of 0.9 nanometers, Nano letters
  19~(11) (2019) 7935--7940.

\bibitem{griesinger2012}
C.~Griesinger, M.~Bennati, H.-M. Vieth, C.~Luchinat, G.~Parigi, P.~H{\"o}fer,
  F.~Engelke, S.~J. Glaser, V.~Denysenkov, T.~F. Prisner, Dynamic nuclear
  polarization at high magnetic fields in liquids., Progress in nuclear
  magnetic resonance spectroscopy 64 (2012) 4--28.

\bibitem{slichter2014}
C.~P. Slichter, The discovery and renaissance of dynamic nuclear polarization,
  Reports on Progress in Physics 77~(7) (2014) 072501.

\bibitem{liu2017}
G.~Liu, M.~Levien, N.~Karschin, G.~Parigi, C.~Luchinat, M.~Bennati,
  One-thousand-fold enhancement of high field liquid nuclear magnetic resonance
  signals at room temperature, Nature chemistry 9~(7) (2017) 676.

\bibitem{plainchont2018}
B.~Plainchont, P.~Berruyer, J.-N. Dumez, S.~Jannin, P.~Giraudeau, Dynamic
  nuclear polarization opens new perspectives for {NMR} spectroscopy in
  analytical chemistry, Analytical Chemistry 90~(6) (2018) 3639--3650.

\bibitem{ardenkjaer2003}
J.~H. Ardenkj{\ae}r-Larsen, B.~Fridlund, A.~Gram, G.~Hansson, L.~Hansson, M.~H.
  Lerche, R.~Servin, M.~Thaning, K.~Golman, Increase in signal-to-noise ratio
  of $>$ 10,000 times in liquid-state {NMR}, Proceedings of the National
  Academy of Sciences 100~(18) (2003) 10158--10163.

\bibitem{capozzi2017}
A.~Capozzi, T.~Cheng, G.~Boero, C.~Roussel, A.~Comment, Thermal annihilation of
  photo-induced radicals following dynamic nuclear polarization to produce
  transportable frozen hyperpolarized $^{13}${C}-substrates, Nature
  communications 8 (2017) 15757.

\bibitem{capozzi2015}
A.~Capozzi, J.-N. Hyacinthe, T.~Cheng, T.~R. Eichhorn, G.~Boero, C.~Roussel,
  J.~J. van~der Klink, A.~Comment, Photoinduced nonpersistent radicals as
  polarizing agents for {X}-nuclei dissolution dynamic nuclear polarization,
  The Journal of Physical Chemistry C 119~(39) (2015) 22632--22639.

\bibitem{kouvril2019}
K.~Kou{\v{r}}il, H.~Kou{\v{r}}ilov{\'a}, S.~Bartram, M.~H. Levitt, B.~Meier,
  Scalable dissolution-dynamic nuclear polarization with rapid transfer of a
  polarized solid, Nature communications 10~(1) (2019) 1--6.

\bibitem{mompean2018}
M.~Mompe{\'a}n, R.~M. S{\'a}nchez-Donoso, A.~De~La~Hoz, V.~Saggiomo, A.~H.
  Velders, M.~V. Gomez, Pushing nuclear magnetic resonance sensitivity limits
  with microfluidics and photo-chemically induced dynamic nuclear polarization,
  Nature communications 9~(1) (2018) 108.

\bibitem{orlando2019}
T.~Orlando, R.~Dervi{\c{s}}o{\u{g}}lu, M.~Levien, I.~Tkach, T.~F. Prisner,
  L.~B. Andreas, V.~P. Denysenkov, M.~Bennati, Dynamic nuclear polarization of
  $^{13}${C} nuclei in the liquid state over a 10 {T} field range, Angewandte
  Chemie 131~(5) (2019) 1416--1420.

\bibitem{eills2019}
J.~Eills, W.~Hale, M.~Sharma, M.~Rossetto, M.~H. Levitt, M.~Utz,
  High-resolution nuclear magnetic resonance spectroscopy with picomole
  sensitivity by hyperpolarization on a chip, Journal of the American Chemical
  Society 141~(25) (2019) 9955--9963.

\bibitem{boero2001}
G.~Boero, J.~Frounchi, B.~Furrer, P.-A. Besse, R.~Popovic, Fully integrated
  probe for proton nuclear magnetic resonance magnetometry, Review of
  Scientific Instruments 72~(6) (2001) 2764--2768.

\bibitem{anders2009}
J.~Anders, G.~Chiaramonte, P.~SanGiorgio, G.~Boero, A single-chip array of
  {NMR} receivers, Journal of Magnetic Resonance 201~(2) (2009) 239--249.

\bibitem{sun2010}
N.~Sun, T.-J. Yoon, H.~Lee, W.~Andress, R.~Weissleder, D.~Ham, Palm {NMR} and
  1-chip {NMR}, IEEE Journal of Solid-State Circuits 46~(1) (2010) 342--352.

\bibitem{anders2011}
J.~Anders, P.~SanGiorgio, G.~Boero, A fully integrated {IQ}-receiver for {NMR}
  microscopy, Journal of Magnetic Resonance 209~(1) (2011) 1--7.

\bibitem{anders2012}
J.~Anders, P.~SanGiorgio, X.~Deligianni, F.~Santini, K.~Scheffler, G.~Boero,
  Integrated active tracking detector for {MRI}-guided interventions, Magnetic
  resonance in medicine 67~(1) (2012) 290--296.

\bibitem{ha2014}
D.~Ha, J.~Paulsen, N.~Sun, Y.-Q. Song, D.~Ham, Scalable {NMR} spectroscopy with
  semiconductor chips, Proceedings of the National Academy of Sciences 111~(33)
  (2014) 11955--11960.

\bibitem{anders2016}
J.~Anders, J.~Handwerker, M.~Ortmanns, G.~Boero, A low-power high-sensitivity
  single-chip receiver for {NMR} microscopy, Journal of Magnetic Resonance 266
  (2016) 41--50.

\bibitem{grisi2015}
M.~Grisi, G.~Gualco, G.~Boero, A broadband single-chip transceiver for
  multi-nuclear {NMR} probes, Review of Scientific Instruments 86~(4) (2015)
  044703.

\bibitem{lei2016}
K.-M. Lei, P.-I. Mak, M.-K. Law, R.~P. Martins, A $\mu${NMR} {CMOS} transceiver
  using a butterfly-coil input for integration with a digital microfluidic
  device inside a portable magnet, IEEE Journal of Solid-State Circuits 51~(10)
  (2016) 2274--2286.

\bibitem{grisi2017}
M.~Grisi, F.~Vincent, B.~Volpe, R.~Guidetti, N.~Harris, A.~Beck, G.~Boero,
  {NMR} spectroscopy of single sub-nl ova with inductive ultra-compact
  single-chip probes, Scientific reports 7 (2017) 44670.

\bibitem{sporrer2017}
B.~Sporrer, L.~Wu, L.~Bettini, C.~Vogt, J.~Reber, J.~Marjanovic, T.~Burger,
  D.~O. Brunner, K.~P. Pruessmann, G.~Tr{\"o}ster, et~al., A fully integrated
  dual-channel on-coil {CMOS} receiver for array coils in 1.5-10.5 {T} {MRI},
  IEEE transactions on biomedical circuits and systems 11~(6) (2017)
  1245--1255.

\bibitem{montinaro2018}
E.~Montinaro, M.~Grisi, M.~Letizia, L.~Peth{\"o}, M.~Gijs, R.~Guidetti,
  J.~Michler, J.~Brugger, G.~Boero, {3D} printed microchannels for sub-nl {NMR}
  spectroscopy, PloS one 13~(5) (2018) e0192780.

\bibitem{grisi2019}
M.~Grisi, G.~M. Conley, P.~Sommer, J.~Tinembart, G.~Boero, A single-chip
  integrated transceiver for high field {NMR} magnetometry, Review of
  Scientific Instruments 90~(1) (2019) 015001.

\bibitem{handwerker2019}
J.~Handwerker, M.~P{\'e}rez-Rodas, M.~Beyerlein, F.~Vincent, A.~Beck,
  N.~Freytag, X.~Yu, R.~Pohmann, J.~Anders, K.~Scheffler, A {CMOS} {NMR} needle
  for probing brain physiology with high spatial and temporal resolution,
  Nature methods (2019) 1--4.

\bibitem{yalcin2008}
T.~Yalcin, G.~Boero, Single-chip detector for electron spin resonance
  spectroscopy, Rev. Sci. Instrum. 79 (2008) 094105.

\bibitem{matheoud2017}
A.~Matheoud, G.~Gualco, M.~Jeong, I.~Zivkovic, J.~Brugger, H.~Rønnow,
  J.~Anders, G.~Boero, Single-chip electron spim resonance detectors operating
  at 50 {GHz}, 92 {GHz}, and 146 {GHz}, J. Magn. Reson. 278 (2017) 113--121.

\bibitem{anders2012b}
J.~Anders, A.~Angerhofer, G.~Boero, K-band single-chip electron spin resonance
  detector, J. Magn. Reson. 217 (2012) 19--26.

\bibitem{gualco2014}
G.~Gualco, J.~Anders, A.~Sienkiewicz, S.~Alberti, L.~Forro, G.~Boero, Cryogenic
  single-chip electron spin resonance detector, J. Magn. Reson. 247 (2014)
  96--103.

\bibitem{matheoud2018}
A.~Matheoud, N.~Sahin, G.~Boero, A single chip electron spin resonance detector
  based on a single high electron mobility transistor, J. Magn. Reson. 294
  (2018) 59--70.

\bibitem{schlecker2018}
B.~Schlecker, A.~Hoffmann, A.~Chu, M.~Ortmanns, K.~Lips, J.~Anders, Towards
  low-cost, high-sensitivity point-of-care diagnostics using {VCO}-based
  {ESR}-on-a-chip detectors, IEEE Sensors Journal 19~(20) (2018) 8995--9003.

\bibitem{poole1996}
C.~P. Poole, Electron spin resonance: a comprehensive treatise on experimental
  techniques, Dover Publications Inc., Mineola, NY, USA, 1996.

\bibitem{goldsborough1960}
J.~Goldsborough, M.~Mandel, G.~Pake, Influence of exchange interaction on
  paramagnetic relaxation times, Physical Review Letters 4~(1) (1960) 13.

\bibitem{gafurov2013}
M.~Gafurov, {TEMPOL} as a polarizing agent for dynamic nuclear polarization of
  aqueous solutions, Magnetic Resonance in Solids. Electronic Journal 15~(1)
  (2013).

\bibitem{petley1984}
B.~Petley, R.~Donaldson, The temperature dependence of the diamagnetic
  shielding correction for proton nmr in water, Metrologia 20~(3) (1984) 81.

\bibitem{hofer2008}
P.~H{\"o}fer, P.~Carl, G.~Guthausen, T.~Prisner, M.~Reese, T.~Carlomagno,
  C.~Griesinger, M.~Bennati, Studies of dynamic nuclear polarization with
  nitroxides in aqueous solution, Applied Magnetic Resonance 34~(3-4) (2008)
  393.

\bibitem{hofer2008b}
P.~H{\"o}fer, G.~Parigi, C.~Luchinat, P.~Carl, G.~Guthausen, M.~Reese,
  T.~Carlomagno, C.~Griesinger, M.~Bennati, Field dependent dynamic nuclear
  polarization with radicals in aqueous solution, Journal of the American
  Chemical Society 130~(11) (2008) 3254--3255.

\bibitem{mccarney2007}
E.~R. McCarney, B.~D. Armstrong, M.~D. Lingwood, S.~Han, Hyperpolarized water
  as an authentic magnetic resonance imaging contrast agent, Proceedings of the
  National Academy of Sciences 104~(6) (2007) 1754--1759.

\bibitem{turke2010}
M.-T. T{\"u}rke, I.~Tkach, M.~Reese, P.~H{\"o}fer, M.~Bennati, Optimization of
  dynamic nuclear polarization experiments in aqueous solution at 15 {MHz}/9.7
  {GHz}: a comparative study with {DNP} at 140 {MHz}/94 {GHz}, Physical
  Chemistry Chemical Physics 12~(22) (2010) 5893--5901.

\bibitem{enkin2014}
N.~Enkin, G.~Liu, I.~Tkach, M.~Bennati, High {DNP} efficiency of {TEMPONE}
  radicals in liquid toluene at low concentrations, Physical Chemistry Chemical
  Physics 16~(19) (2014) 8795--8800.

\bibitem{turke2012}
M.-T. T{\"u}rke, G.~Parigi, C.~Luchinat, M.~Bennati, Overhauser {DNP} with
  $^{15}${N} labelled fr{\'e}my's salt at 0.35 {T}esla, Physical Chemistry
  Chemical Physics 14~(2) (2012) 502--510.

\bibitem{munnemann2008}
K.~M{\"u}nnemann, C.~Bauer, J.~Schmiedeskamp, H.~W. Spiess, W.~Schreiber,
  D.~Hinderberger, A mobile {DNP} polarizer for clinical applications, Applied
  Magnetic Resonance 34~(3-4) (2008) 321--330.

\bibitem{asakura1991}
T.~Asakura, M.~Demura, H.~Ogawa, K.~Matsushita, M.~Imanari, {NMR} imaging of
  diffusion of small organic molecules in silk fibroin gel, Macromolecules
  24~(2) (1991) 620--622.

\bibitem{razavi2011}
B.~Razavi, A 300-{GHz} fundamental oscillator in 65-nm {CMOS} technology, IEEE
  Journal of Solid-State Circuits 46~(4) (2011) 894--903.

\bibitem{takahashi2017}
T.~Takahashi, Y.~Kawano, K.~Makiyama, S.~Shiba, M.~Sato, Y.~Nakasha, N.~Hara,
  Maximum frequency of oscillation of 1.3 {THz} obtained by using an extended
  drain-side recess structure in 75-nm-gate {InAlAs/InGaAs}
  high-electron-mobility transistors, Applied Physics Express 10~(2) (2017)
  024102.

\bibitem{razavi2012}
B.~Razavi, RF microelectronics, 2nd, international ed., Pearson Education
  International, Upper Saddle River, NJ, 2012.

\end{thebibliography}

\end{document}